\documentclass[twocolumn, prb, superscriptaddress]{revtex4-1}
\usepackage{amssymb}
\usepackage{amsmath}
\usepackage{epsfig}
\usepackage{graphics, graphicx}
\usepackage{bbold}
\usepackage{psfrag}
\usepackage{mathcomp}
\usepackage{subfigure}
\usepackage{verbatim}
\usepackage{color}
\usepackage[colorlinks,citecolor=blue]{hyperref}

\graphicspath{{image/}}

\begin{document}
\title{Bridging quantum many-body scar and quantum integrability in Ising chains with transverse and longitudinal fields}
\author{Cheng Peng}
\affiliation{Beijing National Laboratory for Condensed Matter Physics, Institute of Physics, Chinese Academy of Sciences, Beijing 100190, China}
\affiliation{School of Physical Sciences, University of Chinese Academy of Sciences, Beijing 100049, China}
\author{Xiaoling Cui}
\email{xlcui@iphy.ac.cn}
\affiliation{Beijing National Laboratory for Condensed Matter Physics, Institute of Physics, Chinese Academy of Sciences, Beijing 100190, China}
\affiliation{Songshan Lake Materials Laboratory, Dongguan, Guangdong 523808, China}
\date{\today}

\begin{abstract}
Quantum many-body scar (QMBS) and quantum integrability(QI)  have been recognized as two distinct mechanisms for the breakdown of eigenstate thermalization hypothesis(ETH)  in an isolated system. In this work, we reveal a smooth route to connect these two ETH-breaking mechanisms in the Ising chain with transverse and longitudinal fields. Specifically, starting from an initial Ising anti-ferromagnetic state, we find that the  dynamical system undergoes a smooth non-thermal crossover from QMBS to QI by changing the Ising coupling($J$) and longitudinal field($h$) simultaneously while keeping their ratio fixed, which corresponds to the Rydberg Hamiltonian with an arbitrary nearest-neighbor repulsion. Deviating from this ratio, we further identify a continuous thermalization trajectory in ($h,J$) plane that is exactly given by the Ising transition line, signifying an intimate relation between thermalization and quantum critical point.  Finally, we map out a completely different dynamical phase diagram starting from an initial ferromagnetic state, where the thermalization is shown to be equally facilitated by the resonant spin-flip at special ratios of $J$ and $h$. By bridging QMBS and QI in Ising chains, our results demonstrate the breakdown of ETH in much broader physical settings, which also suggest an alternative way to characterize quantum phase transition via thermalization in non-equilibrium dynamics.
\end{abstract}
\maketitle
 
\section{Introduction}
Non-equilibrium dynamics of a quantum ergodic system can end at thermalization  at long time, where the local observables show thermal equilibrations 
that are well described by the standard statistical mechanics. 
The underlying mechanism of this phenomenon is the eigenstate thermalization hypothesis(ETH)
\cite{Deutsch1991,Srednicki1994,Olshanii2008,Huse2014,Beugeling2014,Rigol2016review,Deutsch2018review}, which views  each eigenstate as a thermal state in the sense that a local subsystem can be thermalized by the rest of whole system.  Taking different initial states, the thermalization can be further classified to weak or strong type, where the local observables converge to thermal values completely or after time average\cite{Banuls2011weak,Motrunich2017weak, Liu2019, Fan2021}. 

However, not all quantum systems obey the ETH and thermalize. Typical examples violating the ETH include  1D integrable models with numerous conservation laws\cite{Weiss2006QI,Rigol2007QI,Calabrese2011QI,Essler2016QI,Rigol2016GGE}, and the many-body localization in the presence of strong disorder\cite{Basko2006MBL,Serbyn2013MBL,Huse2014MBL,RMP2019MBL}. Besides,  a new class of ETH-breaking dynamics  has been reported recently in the quantum simulator of Rydberg atoms\cite{Lukin2017Rydberg} taking an initial Ising anti-ferromagnetic state $|Z_2\rangle=|\downarrow \uparrow ...\downarrow\uparrow\rangle$, with $\uparrow$ (or $\downarrow$) representing the Rydberg (or atomic ground) state. Later it was shown that the absence of thermalization there can be attributed to the weak ergodicity breaking due to non-thermal eigenstates embedded in the sea of thermal eigenstates, dubbed the ``quantum many-body scar"(QMBS)\cite{Papic2018ScarNP,Papic2018scarPRB,Ho2019scar,Papic2021scarReview, review_Bernevig, review_Moessner}. A celebrated model for QMBS is the PXP Hamiltonian (see Eq.\ref{pxp}), which assumes a hard-core repulsion between Rydberg atoms at neighboring sites and thus projects out all configurations with two adjacent $\uparrow$.  The presence of such constraint indicates a close relation between QMBS and confinement theory as well as lattice gauge models\cite{Papic2021scarReview}. 
Apart from PXP, various other models hosting QMBS have been pointed out theoretically\cite{Bernevig2018ScarHubbard,Bernevig2020ScarAKLT,Motrunich2020ScarAKLT,Bernevig2020ScarHubbard,Motrunich2020ScarHubbard,SchecterScarXY,Zhao2020,Zhao2021}.  Very recently, QMBS was also observed in the tilted Bose Hubbard setup that realizes the PXP model with an additional magnetic field\cite{Yuan2022Z0scar}, while the initial state is not $|Z_2\rangle$ but the ferromagnetic state $|Z_0\rangle=|\downarrow \downarrow ...\downarrow\downarrow\rangle$. 

The newly uncovered QMBS suggests that it is still far from a full understanding of thermalization in isolated quantum systems. In particular, how robust is QMBS when the parameter constraint (as in PXP) is relaxed in a more general and realistic situation(such as disorder\cite{Martin})? and whether there is any connection between QMBS and other ETH-breaking mechanisms such as QI?   
Moreover, given the important role of initial state played in QMBS, it is worthwhile to explore the initial-state-dependence of thermalization behavior in above general situations. 
These studies will help to understand QMBS in a much broader physical context, and even more, to bridge the gap between different ETH-breaking phenomena discovered so far.

\begin{widetext}

\begin{figure}[h]
\includegraphics[height=6cm]{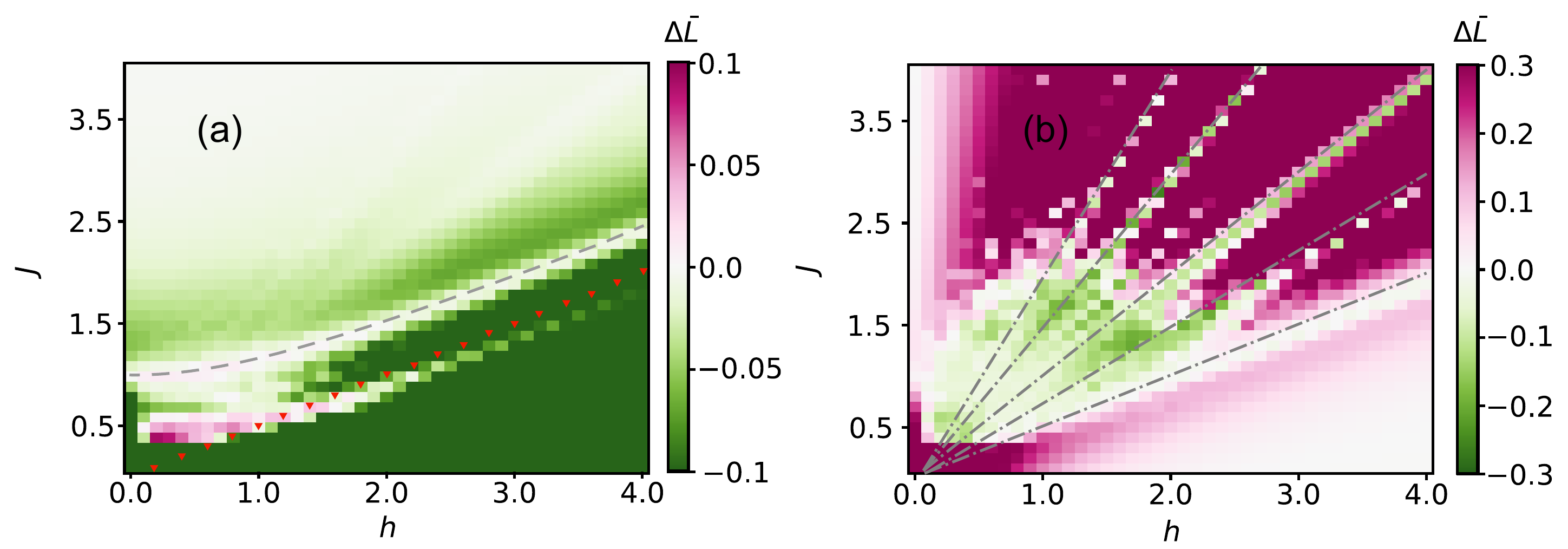}
\caption{Dynamical phase diagram in $(h,J)$ plane for the ICTLF system starting from an initial $|Z_2\rangle$ (a) or $|Z_0\rangle$ (b) state. The color shows the difference $\Delta \bar{L}$ (Eq.\ref{diff}) for an infinite chain, and  the green or red color with visible $|\Delta \bar{L}|$ corresponds to non-thermalization. 
In (a), gray dashed curve denotes the  Ising transition point, and red dotted line is given by $J/h=1/2$ that describes the Rydberg system with arbitrary coupling. In (b), gray dash-dot lines mark the location of resonant spin flip as given by $J/h=2, \ 3/2,\ 1,\ 3/4,\ 1/2$ (from top to bottom).  }\label{diagram}
\end{figure}

\end{widetext}

In this work, we examine the robustness of QMBS as well as its connection with QI in the Ising chains with transverse and longitudinal fields (ICTLF). The model of ICTLF has been realized for interacting bosons in tilted optical lattice\cite{Greiner2011IsingExps} following its original proposal\cite{Sachdev2002}.  By varying the strength of Ising coupling ($J$) and longitudinal field ($h$), ICTLF can well cover the parameter regimes of both QMBS and QI, and therefore provides an ideal platform for the study. 
In particular, by changing $J$ and $h$ simultaneously while keeping their ratio fixed ($J/h=1/2$), this model reproduces  Rydberg Hamiltonian with an arbitrary nearest-neighbor repulsion. In this case, we find that the $|Z_2\rangle$-initiated dynamical system  undergoes a smooth non-thermal crossover from QMBS to QI.  During the process, QMBS physics is quite robust and it can even persist for finite Rydberg repulsions that are just a few times of the transverse field. When deviating from this ratio ($J/h\neq 1/2$), we identify a continuous trajectory in the ($h,J$) plane where the local observables tend to thermalize, and such trajectory turns out to exactly match the Ising transition line (see Fig.\ref{diagram}(a)). This remarkable consistence signifies an intimate relation between thermalization and quantum phase transition(QPT), as also discussed earlier in PXP model\cite{Zhai2022}  that corresponds to a special limit of our system.  
Here we attribute this phenomenon to the enhanced quantum ergodicity by large correlation length of low-energy states near QPT.  Finally, we map out a completely different dynamical phase diagram for the initial $|Z_0\rangle$ state(see Fig.\ref{diagram}(b)), where the thermalization is shown to be equally favored by the resonant spin-flip at special ratios of $J$ and $h$. These results can be readily tested in Rydberg systems or bosons in tilted lattices.


The rest of the paper is organized as follows. In section \ref{sec_II}, we present the model Hamiltonian and the formula of non-equilibrium dynamics. The thermalization properties starting from the initial $|Z_2\rangle$ and $|Z_0\rangle$ are discussed, respectively, in section \ref{sec_III} and \ref{sec_IV}. The final section \ref{sec_V} is contributed to the summary of our results.  In appendix \ref{appendix_a} and  \ref{appendix_b}, we show the robustness of our results against the choice of momentum interval in numerical calculation and against the finite-size effect.

\section{Model and Formalism} \label{sec_II}

In this section, we present the model Hamiltonian studied in this work, as well as the formalism for non-equilibrium dynamics. 
 
\subsection{ICTLF and its reduction to various models}

We first write down the model Hamiltonian for an Ising chain with transverse and longitudinal fields (ICTLF):
\begin{equation}
H_{\rm ICTLF}(J,h)=\sum_i \sigma_{i}^x + J\sum_i \sigma_i^z\sigma_{i+1}^z + h\sum_i \sigma_{i}^z. \label{Ising}
\end{equation}
Here $\sigma_i^{\alpha}$($\alpha=x,y,z$) is the Pauli matrix at site-$i$. Throughout the paper,  we take the energy unit as the transverse field ($=1$), and thus the Ising coupling $J$ and longitudinal field $h$ are all dimensionless parameters. 

When $J=h/2$, the last two terms in (\ref{Ising}) reduce to the nearest-neighbor interaction between spin-$\uparrow$ atoms, i.e., $\sum_i n_{i\uparrow}n_{i+1\uparrow}$ with $n_{i\uparrow}=(1+\sigma_i^z)/2$. In this case, (\ref{Ising}) can describe the Rydberg atoms($\uparrow$) in optical lattices with nearest-neighbor interaction $V\equiv4J=2h$:
\begin{equation}
H_{\rm Ryd}(V)=\sum_i \sigma_{i}^x + V\sum_i n_{i,\uparrow}n_{i+1,\uparrow}. \label{H_V}
\end{equation}
Further sending $V\rightarrow\infty$, (\ref{H_V}) gives the PXP model:
\begin{equation}
H_{\rm PXP}=\sum_i P_{i-1,\downarrow} \sigma_{i}^x P_{i+1,\downarrow}, \label{pxp}
\end{equation}
where $P_{i,\downarrow}=(1-\sigma^z_i)/2$ is the projection operator that only allows the occupation of spin-$\downarrow$ at site-$i$. Under (\ref{pxp}), the spin-flip at any site can only occur when its neighboring spins are not $\uparrow$, thereby excluding the situation when $\uparrow$ atoms are at neighboring sites, as expected for the hard-core interaction $V\rightarrow\infty$. 

\begin{figure}[t]
\includegraphics[width=8cm]{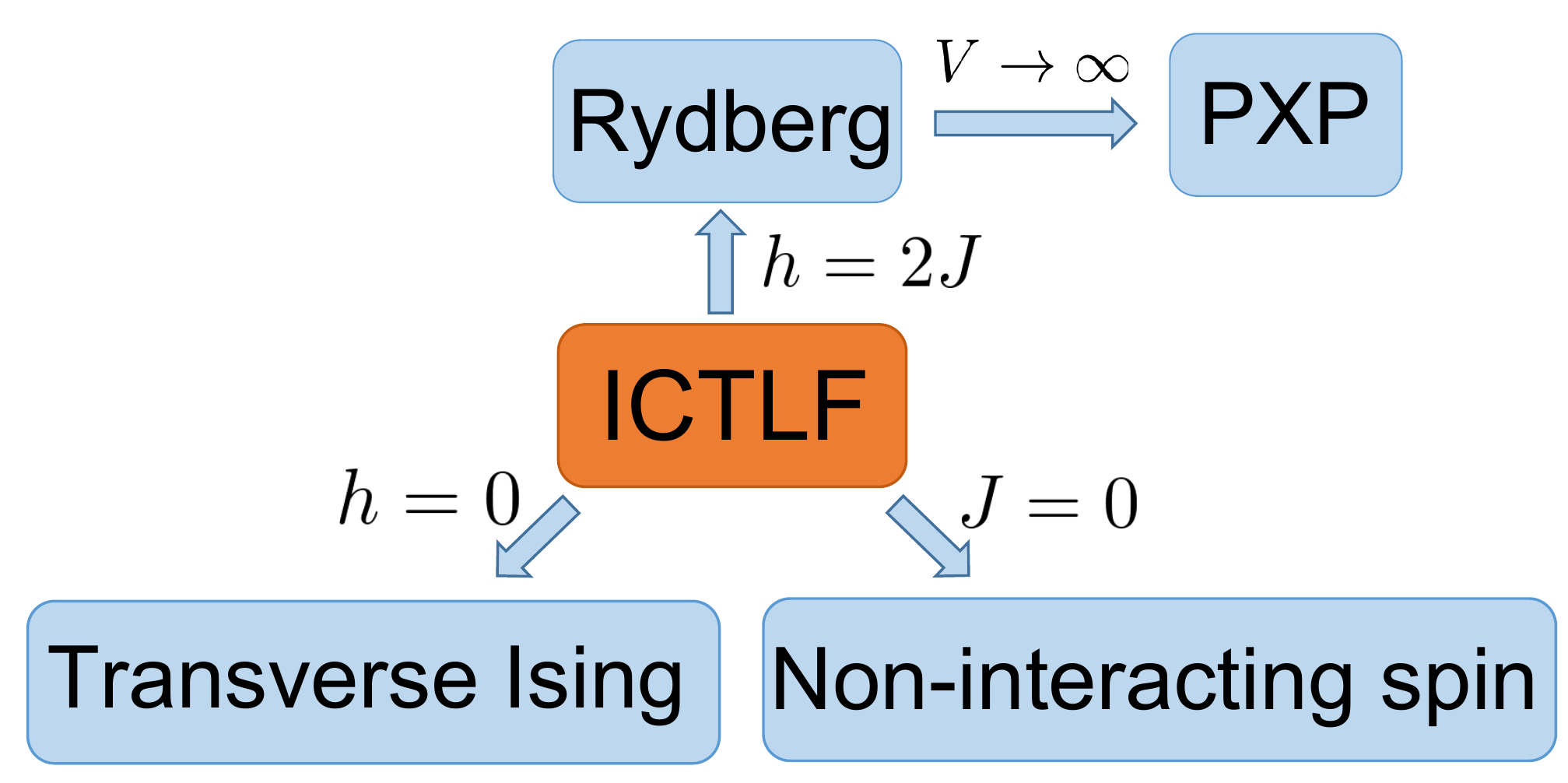}
\caption{Reduction of the Ising chain with transverse and longitudinal fields (ICTLF) to various models by varying $J$ and $h$.  }
\label{models}
\end{figure}

When $h=0$,  ICTLF (\ref{Ising}) reduces to the integrable model of a transverse Ising chain, which is exactly solvable. By sending $J\rightarrow 0$, (\ref{Ising}) can also cover the non-interacting limit of spin model. As summarized  in Fig.\ref{models}, by varying $J$ and $h$,  ICTLF can reproduce various models that were previously studied in different contexts, including the Rydberg model (with PXP as  a special limit), integrable model of transverse Ising, and non-interacting spin model. In this way, ICTLF serves an ideal platform to bridge different systems and regimes with different thermalization properties.

\subsection{Formalism for non-equilibrium dynamics}

The time evolution of the system described by  Hamiltonian $\hat{H}$ and starting from  initial state $\psi(t=0)$ can be obtained as:
\begin{equation}
|\psi(t)\rangle = e^{-i\hat{H}t}|\psi(t=0)\rangle = \sum_{\alpha} e^{-i E_\alpha t} \langle\alpha|\psi(t=0)\rangle |\alpha\rangle,
\end{equation}
where $|\alpha\rangle$ denotes the eigenstate of $\hat{H}$ and $E_{\alpha}$ is its eigenenergy. In this work we take the periodic boundary condition and focus on the expectation value of local operator 
\begin{equation}
\hat{L} = \sigma_1^z\sigma_2^z, 
\label{L}
\end{equation}
At sufficiently long time, the average value of $\langle \hat{L} \rangle$ is given by 
\begin{equation}
\bar{L} =  \sum_{\alpha} \langle\alpha|\hat{L}|\alpha\rangle |\langle\alpha|\psi(t=0)\rangle|^2,  \label{L_infty}
\end{equation}
where we have only kept the diagonal ensemble average and dropped the off-diagonal ones due to long time average. 
In our numerics,  we have exactly diagonalized the lattice system for different sizes\cite{quspin1,quspin2}  and obtained the time evolution $L(t)\equiv \langle \psi(t)|\hat{L}|\psi(t) \rangle$ as well as its long-time average $\bar{L}$ (Eq.\ref{L_infty}).

On the other hand, we use the Gibbs ensemble to calculate the thermal prediction  of $\bar{L}$. Given the total energy and momentum are both conserved by the Hamiltonian, we define the thermal density matrix and distribution function as:
\begin{align}
&\hat{\rho}_{th} = \frac{1}{Z}e^{-\beta\hat{H}-\lambda\hat{K}}, \\
&Z = Tr(e^{-\beta\hat{H}-\lambda\hat{K}}).
\end{align}
Here the thermodynamic parameters $\beta$ and $\lambda$ are determined by:
\begin{equation}
	\bar{E} = Tr(\hat{\rho}_{th}\hat{H} ),\ \ \ \ 	\bar{K} = Tr(\hat{\rho}_{th}\hat{K}),
\end{equation}
with $\bar{E}$ and $\bar{K}$ fixed by the initial condition:
\begin{equation}
\begin{split}
 \bar{E} &= \langle\psi(t=0)|\hat{H}|\psi(t=0)\rangle, \\
 \bar{K} &= \langle\psi(t=0)|\hat{K}|\psi(t=0)\rangle.
\end{split}
\end{equation}
Once we obtain $\beta$ and $\lambda$, the thermal prediction of $\bar{L}$ is given by:
\begin{equation}
	\bar{L}_{th} = Tr(\hat{\rho}_{th}\hat{L}).
\end{equation}
In this work, we will consider the momentum interval as $[0,2\pi)$. ln appendix \ref{appendix_a}, we show that the other choice of momentum interval,  such as $(-\pi,\pi]$, will not qualitatively change the results. 

Finally, one can  examine the difference between  $\bar{L}$ from exact diagonalization and $\bar{L}_{th}$ from thermal prediction:
\begin{equation}
\Delta \bar{L}= \bar{L}-\bar{L}_{th}, \label{diff}
\end{equation}
 which can serve as an important quantity for the thermalization property of the system. 
Namely, a vanishingly small $\Delta \bar{L}$ is a necessary, while not sufficient, condition for thermalization. In other words, the thermalization requires $\Delta \bar{L}\rightarrow 0$, which means that  if $|\Delta \bar{L}|$ is visibly large then one can definitely conclude non-thermalization of the system. 

 In our numerical calculations, we have exactly diagonalized the Hamiltonians  (\ref{Ising},\ref{H_V}) for different lattice sizes $N_L$ and obtained the long time average $\bar{L}$. For comparison, we also calculate the thermal average $\bar{L}_{th}$ for various $N_L$. Due to heavy numerics involved for large lattice size, we have taken the cutoff of $N_L$ as $18$ in computing $\bar{L}$ and as $16$ in  $\bar{L}_{th}$. The finite-size scaling is then carried out to extract the results for an infinite chain $1/N_L\rightarrow 0$, which give rise to the phase diagrams shown in Fig.\ref{diagram}. We have checked that the main features of these diagrams, as discussed in later sections, do not rely on the scaling but robustly show up for a finite system, see discussions in appendix \ref{appendix_b}. 

\section{Thermalization property starting from $|Z_2\rangle$} \label{sec_III}

In this section, we show the results for the long-time dynamics of ICTLF system  starting from $|Z_2\rangle$ state. The according phase diagram in ($h,J$) plane is presented in Fig.\ref{diagram}(a). We will mainly focus on two lines on the diagram: one is along $J/h=1/2$ (red dotted line) corresponding to Rydberg system with arbitrary nearest-neighbor interaction, which displays non-thermalization all along;  the other is along the Ising transition line (gray dashed line) which displays strong tendency towards  thermalization.  Together with the two phenomena, a sequence of smooth crossovers between non-thermal and thermal regimes will be analyzed as changing $J$ in large $h$ limit.

\subsection{Non-thermal crossover along $J/h=1/2$}

We first consider the line $J/h=1/2$ in Fig.\ref{diagram}(a), which corresponds to the Rydberg system with tunable nearest-neighbor interaction $V\equiv 4J=2h$, see (\ref{H_V}). When $V\rightarrow 0$, the Rydberg model (\ref{H_V}) reduces to non-interacting limit that is also a special case of QI; when $V\rightarrow \infty$, (\ref{H_V}) reduces to the PXP Hamiltonian (\ref{pxp})  that hosts QMBS. Therefore, by increasing $V$ from zero to large, the system will evolve from QI to QMBS, and we will show below that such evolution is a smooth crossover without thermalization all along.


\begin{figure}[h]
\includegraphics[width=6cm, height=8cm]{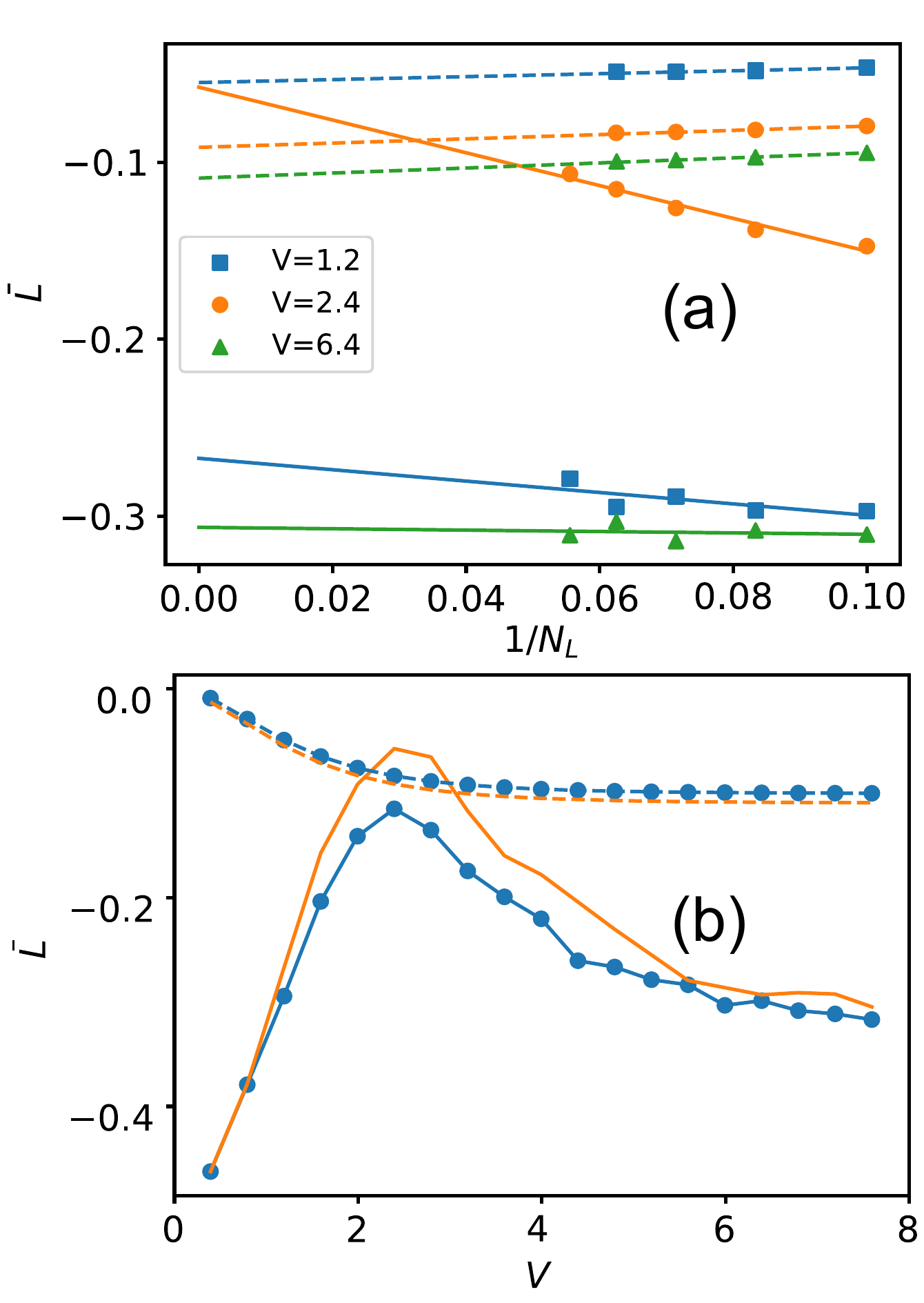}
\caption{ Non-thermal crossover for Rydberg system with tunable interaction $V$ and starting from $|Z_2\rangle$ state. (a)Finite-size scaling to $\bar{L}$ from exact diagonalization (solid lines) and $\bar{L}_{th}$ from thermal prediction (dashed). The squares, circles and triangles are for  $V=1.2,\ 2.4$ and $6.4$ respectively. $N_L$ is the number of lattice site. (b) $\bar{L}$ (solid lines) and $\bar{L}_{th}$ (dashed) as functions of $V$ for $N_L=16$ (with circle) and for an infinite chain (no circle). }
\label{fig_finite_V}
\end{figure}


 In Fig.\ref{fig_finite_V}(a), we show the long time average $\bar{L}$ and thermal average $\bar{L}_{th}$ for various $N_L$ at three typical $V$, where the finite-size scaling is carried out for both quantities. Finally, we plot out  $\bar{L}$ and $\bar{L}_{th}$ as functions of $V$ in Fig.\ref{fig_finite_V}(b). We can see that $\bar{L}$ evolves non-monotonically as increasing $V$, in contrast to the monotonic decrease of $\bar{L}_{th}$. In the small $V$ regime,  $\bar{L}$ increases linearly with $V$ and it reaches maximum  near $V\sim 2.5$. Further increasing $V$,  $\bar{L}$ starts to decrease and finally saturates at certain finite value when $V\rightarrow \infty$, i.e., the PXP limit. During this process, $\bar{L}$ shows visible deviation from the thermal prediction $\bar{L}_{th}$, except for two fine-tuning points when the two values cross. Therefore, one can conclude a non-thermal crossover for the Rydberg system from non-interacting to PXP limit all along. 

\begin{figure}[h]
\includegraphics[width=8.5cm]{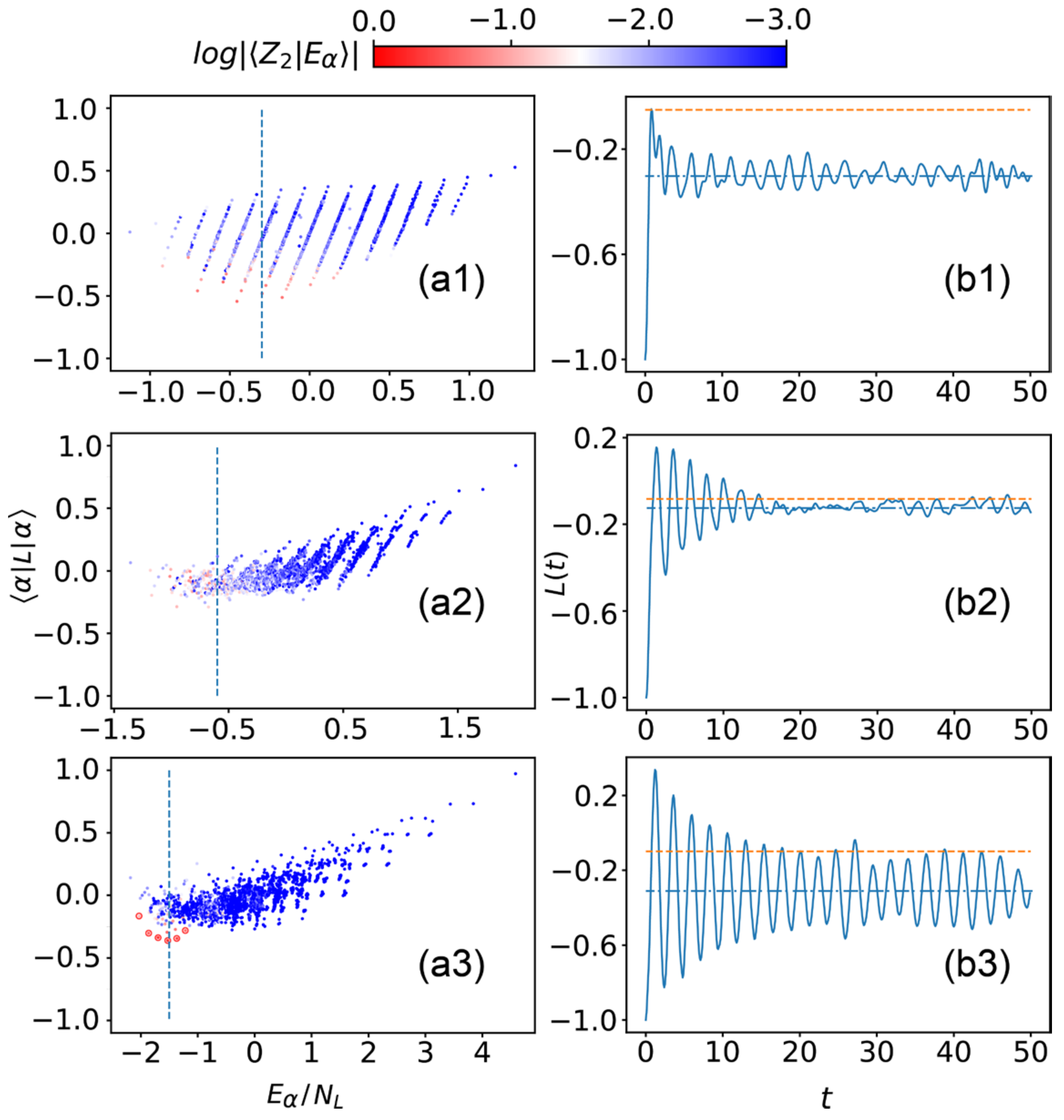}
\caption{Diagonal elements of $\hat{L}$ in energy eigenbasis (a1,a2,a3) and time evolution $L(t)$ (b1,b2,b3) during the non-thermal crossover of Rydberg atoms starting from $|Z_2\rangle$. Here we take three Rydberg couplings  $V=1.2$(a1,b1), $2.4$(a2,b2) and $6$(a3,b3), and the lattice size is $N_L=16$.  In (a1,a2,a3), the color shows the overlap between each eigenstate and $|Z_2\rangle$, and the vertical dashed line denotes the location of initial energy $E(t=0)$. The scar states in (a3) are highlighted by red circles. In (b1,b2,b3), the red dashed line shows the thermal prediction $\bar{L}_{th}$, and the blue dot-dashed  line in  shows $\bar{L}$.}\label{fig_finite_V_2}
\end{figure}

To gain more understanding on the non-thermal crossover, we have taken three typical values of $V$ and examined the diagonal elements of $\hat{L}$ in energy eigenbasis as well as the actual time evolution $L(t)$ in Fig.\ref{fig_finite_V_2}.  
For small $V$ that is close to the QI (non-interacting) limit, the distribution $\langle L \rangle_{\alpha}\sim E_{\alpha}$  appears as stripes but does not converge into narrow curve as in typical ETH scenario, see Fig.\ref{fig_finite_V_2}(a1). Accordingly, $L(t)$ shows non-thermal dynamics in Fig.\ref{fig_finite_V_2}(b1), where the long-time average visibly deviates from the thermal prediction (horizontal line). Further increase $V$ as in Fig.\ref{fig_finite_V_2}(a2,b2), the stripes become smeared out in low-energy space, while the distribution of $\langle L \rangle $ at $E\sim E(t=0)$ is still quite expanded. 
As a result, the long-time average of $L(t)$ still deviates from $\bar{L}_{th}$.   For an even larger $V$, the physics of QMBS starts to emerge. As shown in Fig.\ref{fig_finite_V_2}(a3,b3), at $V=6$ there exists a set of discrete energy levels with nearly equal energy spacing and with the largest overlap with $|Z_2\rangle$; accordingly,  $L(t)$ shows a visible and periodic oscillation even at long time and its average cannot be determined by the thermal value.  

In above we have seen a smooth crossover of the non-thermal dynamics from QI (small $V$) to QMBS (large $V$) regimes,  taking $|Z_2\rangle$ as initial state.  The location of crossover is roughly given by $V\sim 2.5$, when $\bar{L}$ reaches maximum and well separates the two regimes (see Fig.\ref{fig_finite_V}(b)). Interestingly, we can see that the key features of QMBS already emerge at finite $V$ that is just a few times of transverse field. This liberates QMBS from the extreme PXP limit and extends it to a much more realistic and broader parameter regime.


\subsection{Thermalization facilitated  by quantum phase transition} 

When $J/h\neq1/2$, we find a remarkable feature for the $|Z_2\rangle$-initiated dynamics, i.e., the tendency of thermalization at the critical point of quantum phase transition (QPT). As shown in Fig.\ref{diagram}(a), the gray dashed line  marks the location of Ising transition from paramagnetic to anti-ferromagnetic phases\cite{Landau, Ovchinnikov}, which smoothly connects the QI regime at $h=0$ to QMBS regime at large $h$. We will show below that the dynamical system tends to thermalize exactly on this line. 

\begin{figure}[h]
\includegraphics[width=8.5cm, height=8cm]{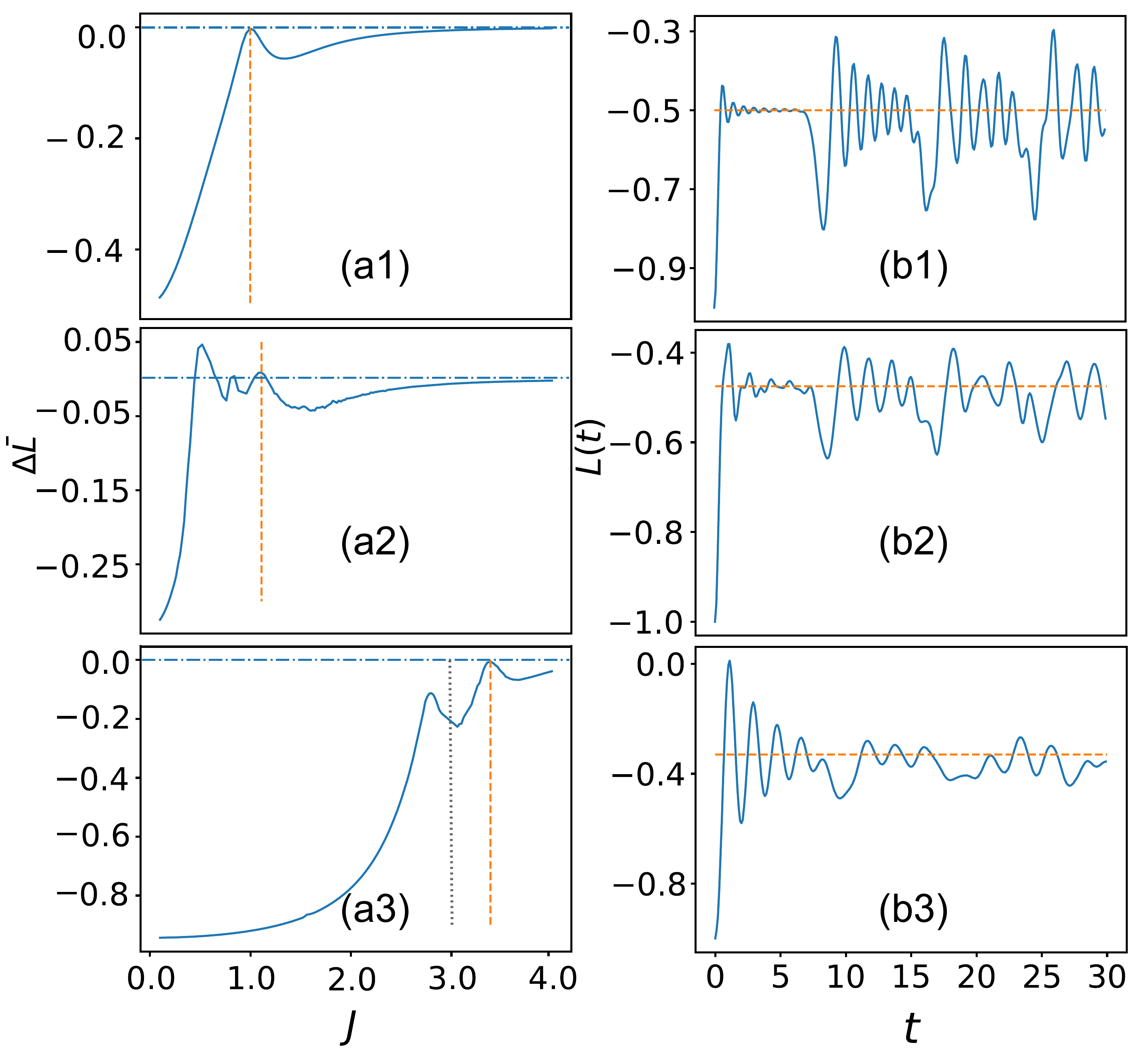}
\caption{Thermalization of $|Z_2\rangle$-initiated dynamics at the Ising transition point. (a1,a2,a3)  $\Delta \bar{L}$ as a function of $J$ for fixed $h=0$(a1), $0.8$(a2) and $6$(a3). The vertical dashed orange lines mark the location of Ising transition: $J_c=1$(a1), $1.11$(a2) and $3.4$(a3).  Vertical dotted gray line in (a3) mark the location of QMBS with $J=h/2=3$. The horizontal dash-dot lines show $\Delta \bar{L}=0$. (b1,b2,b3) $L(t)$ at the Ising transition points as located  in (a1,a2,a3).  Red dashed lines show the thermal predictions $\bar{L}_{th}$.  Here we take lattice size $N_L=16$. 
}
\label{fig_QPT}
\end{figure}

In Fig.\ref{fig_QPT}(a1,a2,a3), we take three different values of $h$ and plot out  $\Delta \bar{L}$ as a function of $J$. We can see that all $\Delta \bar{L}\sim J$ share similar non-monotonic lineshape, and all $\Delta \bar{L}$ drop to nearly zero  at certain $J=J_c$  signifying the tendency of thermalization. Remarkably, we find that the thermalization location $J_c$ exactly matches the Ising transition point, as denoted by vertical dashed lines in Fig.\ref{fig_QPT}(a1,a2,a3). Here we have extracted the Ising transition points independently through the universal scaling of lowest energy gap near the transition\cite{Sachdev2002}. In Fig.\ref{fig_QPT}(b1,b2,b3), we further show the dynamics of $ L(t)$ at the three critical points in (a1,a2,a3). We can see that  the thermal value $\bar{L}_{th}$ indeed matches the long-time average $\bar{L}$ in each case. However, the dynamics $L(t)$ itself at long time may still oscillate heavily, especially for the integrable case of  $h=0$ (see Fig.\ref{fig_QPT}(b1)). This is also consistent with a general belief that  an integrable system does not thermalize. 
As increasing $h$ to enter the non-integrable regime (Fig.\ref{fig_QPT}(b2,b3)), we can see that the oscillation amplitude becomes gradually smaller, and in this case we can call it as a weak thermalization.

 In fact, a negligible $\Delta \bar{L}$ also shows up at large $J(\gg 1)$ even for the integrable case ($h=0$), as shown by the white color region in Fig.\ref{diagram}(a). In this limit the Ising term dominates the Hamiltonian and the initial state $|Z_2\rangle$ can well approximate the ground state. This is why the time evolution of the system displays very weak dynamics around this initial state. Similar situation also applies for the $|Z_0\rangle$-initiated dynamics in the regime $J\gg 1$ at small $h$, see Fig.\ref{diagram}(b), where $|Z_0\rangle$ is a good approximation for the highest eigenstate of the system instead of the lowest one.

\begin{figure}[h]
\includegraphics[width=9cm,  height=8cm]{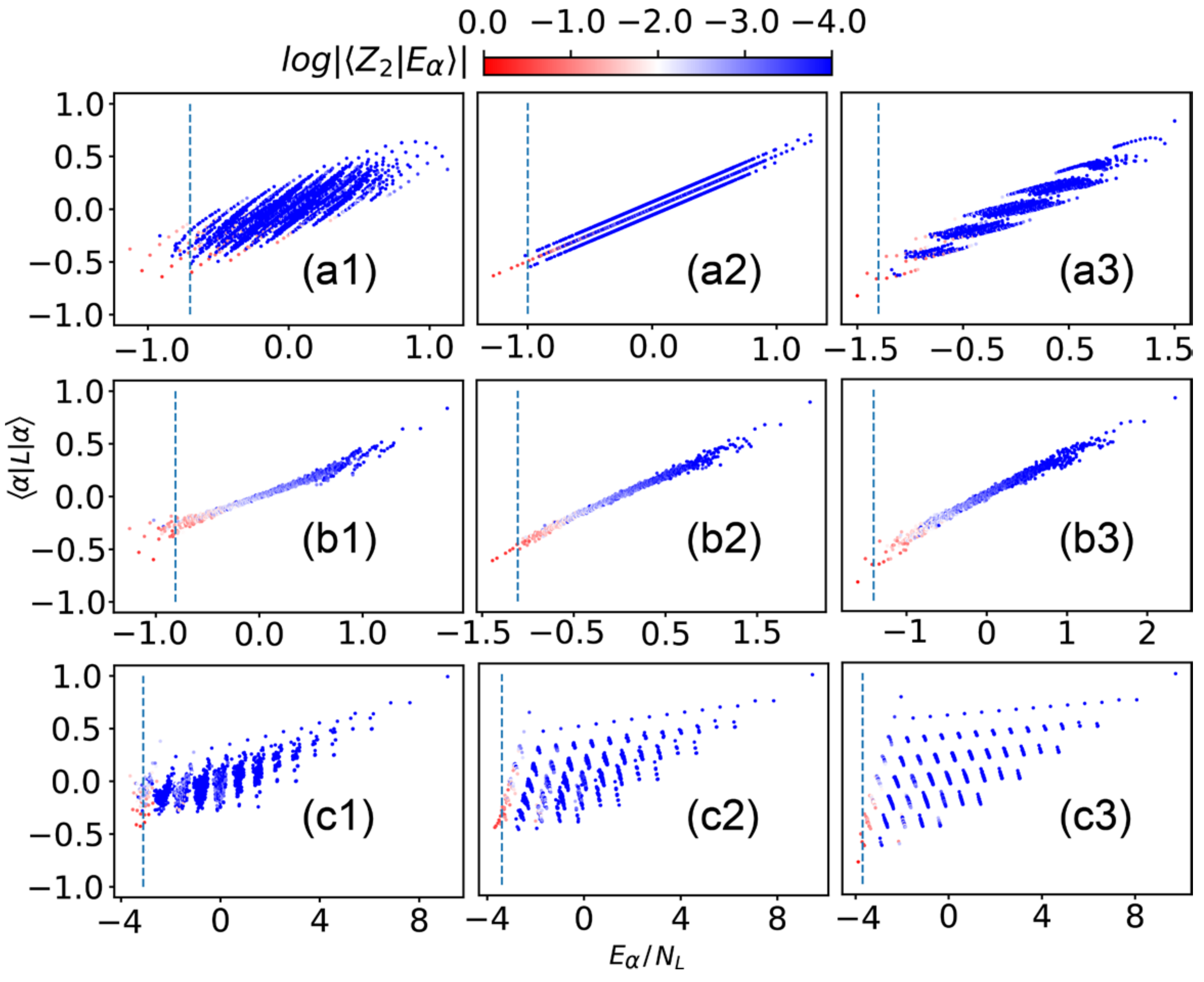}
\caption{Diagonal elements of $\hat{L}$ in energy eigenbasis for different $(h,J)$ parameters near the Ising transition line. In (a1,a2,a3), we have $h=0$ and $J=0.7$(a1), $1(=J_c)$(a2) and $1.3$(a3). In (b1,b2,b3), we have $h=0.8$ and  $J=0.81$(b1), $1.11(=J_c)$(b2) and $1.41$(b3).  In (c1,c2,c3),  we have $h=6$ and  $J=3.1$(c1), $3.4(=J_c)$(c2) and $3.7$(c3). The lattice size is $N_L=16$. In all plots, the color shows the overlap between each eigenstate and initial $|Z_2\rangle$ state, and the vertical dashed line marks the location of initial state energy $E(t=0)$. }
\label{fig_QPT2}
\end{figure}

To explore the intrinsic relation between thermalization and QPT, in Fig.\ref{fig_QPT2} we plot out the diagonal elements of  $\hat{L}$ in energy eigenbasis near (below, at and above) Ising transition points, where the color denotes the overlap of each eigenstate with initial $|Z_2\rangle$ state. For the integrable case at $h=0$, at the transition point  $J=J_c=1$ as shown in Fig.\ref{fig_QPT2}(a2), the $\langle L \rangle_{\alpha}\sim E_{\alpha}$ dependence fall into three straight lines that are very close to each other, and the largest overlap with $|Z_2\rangle$ appears solely within the central line. Accordingly, the resulted dynamics of $L(t)$ shows strong oscillations with long-time average matching the thermal prediction (see Fig.\ref{fig_QPT}(b1)). In comparison, when $J<J_c$ (Fig.\ref{fig_QPT2}(a1)) or $J>J_c$(Fig.\ref{fig_QPT2}(a3)), the distribution of $\langle L \rangle_{\alpha}\sim E_{\alpha}$  no longer falls into narrow lines but is rather expanded. Such non-thermal distribution leads to a visible deviation between $\bar{L}$ and $\bar{L}_{th}$ (see also $\Delta \bar{L}\sim J$ in Fig.\ref{fig_QPT}(a1)). For non-integrable cases at finite $h$, similar features of $\langle L \rangle_{\alpha}\sim E_{\alpha}$ near $J_c$ also show up (see Fig.\ref{fig_QPT2}(b1,b2,b3) and (c1,c2,c3)). 

It is noted that the thermalization enhanced by QPT has also been pointed out in Ref.\cite{Zhai2022} based on the PXP model with an additional magnetic field $m$. Our system can effectively reproduce this model with $m=h-2J$ when considering the limit $h,J\gg 1$ and $h\sim 2J$. However, a crucial difference should be noted. In the magnetic PXP model, the thermal property of $\langle \hat{L}\rangle$ (or any other local  observable that is a product of $\sigma^z_i$) starting from $|Z_2\rangle$ or $|Z_0\rangle$ state is unchanged  under  the transformation $m\rightarrow -m$. This means that the thermalization should appear for both $m>0$ and $m<0$ sides with the same $|m_c|$. In contrast, in our ICTLF model we have not found such symmetry, see Fig.\ref{diagram}(a). This can be attributed to the different Hilbert spaces associated with these two models. In particular, in $h>2J$ (or $m>0$) side, the critical state responsible for the phase transition is at high energy but not the ground state, and the initial energy of $|Z_2\rangle$ is also quite high.  In this case, the thermodynamic parameter $\beta$  in the magnetic PXP model is negative, while in our ICTLF model is always positive. 
This difference is a direct consequence of the much larger Hilbert space in ICTLF  than in magnetic PXP, which can greatly affect the thermal property initiated from high energy states.





In above we have demonstrated that the thermalization right at the Ising transition point  is closely related to the convergent distribution of local operators for relevant energy states. Physically, this phenomenon can be attributed to the large correlation length at the transition point, which leads to strong hybridization  between different states and therefore enhances the quantum ergodicity of the system. On the other hand, we should note that the QPT refers to the change of ground state, and the large correlation length near QPT only applies to low-energy states but not those high ones. As a result, the enhanced ergodicity requires the initial state staying in the low-energy space, and this is obviously satisfied by $|Z_2\rangle$ state which has a low initial energy all along the transition line (see vertical lines in Fig.\ref{fig_QPT2}(a2,b2,c2)).  As we will show in the next section, if change the initial state to be in the high-energy manifold, such as $|Z_0\rangle$ state, the QPT-enhanced thermalization will no longer hold true. 



\subsection{Non-thermal to thermal crossovers as varying $J$ for a large $h$}

In above two subsections, we have investigated the non-thermal crossover along $J/h=1/2$ and the quantum ergodicity enhanced by Ising transitions for the $|Z_2\rangle$-initiated dynamics.  Together with these two phenomena, we can see 
a sequence  of smooth crossovers between non-thermal and thermal regimes as increasing $J$ for a large $h$ in Fig.\ref{diagram}(a). These crossovers are clearly reflected by the non-monotonic evolution of $\Delta \bar{L}\sim J$ as shown in Fig.\ref{fig_QPT}(a3). In the following, we will take a fixed $h=6$ and a tunable $J$ to analyze in detail these crossovers.   

\begin{figure}[h]
\includegraphics[width=9cm]{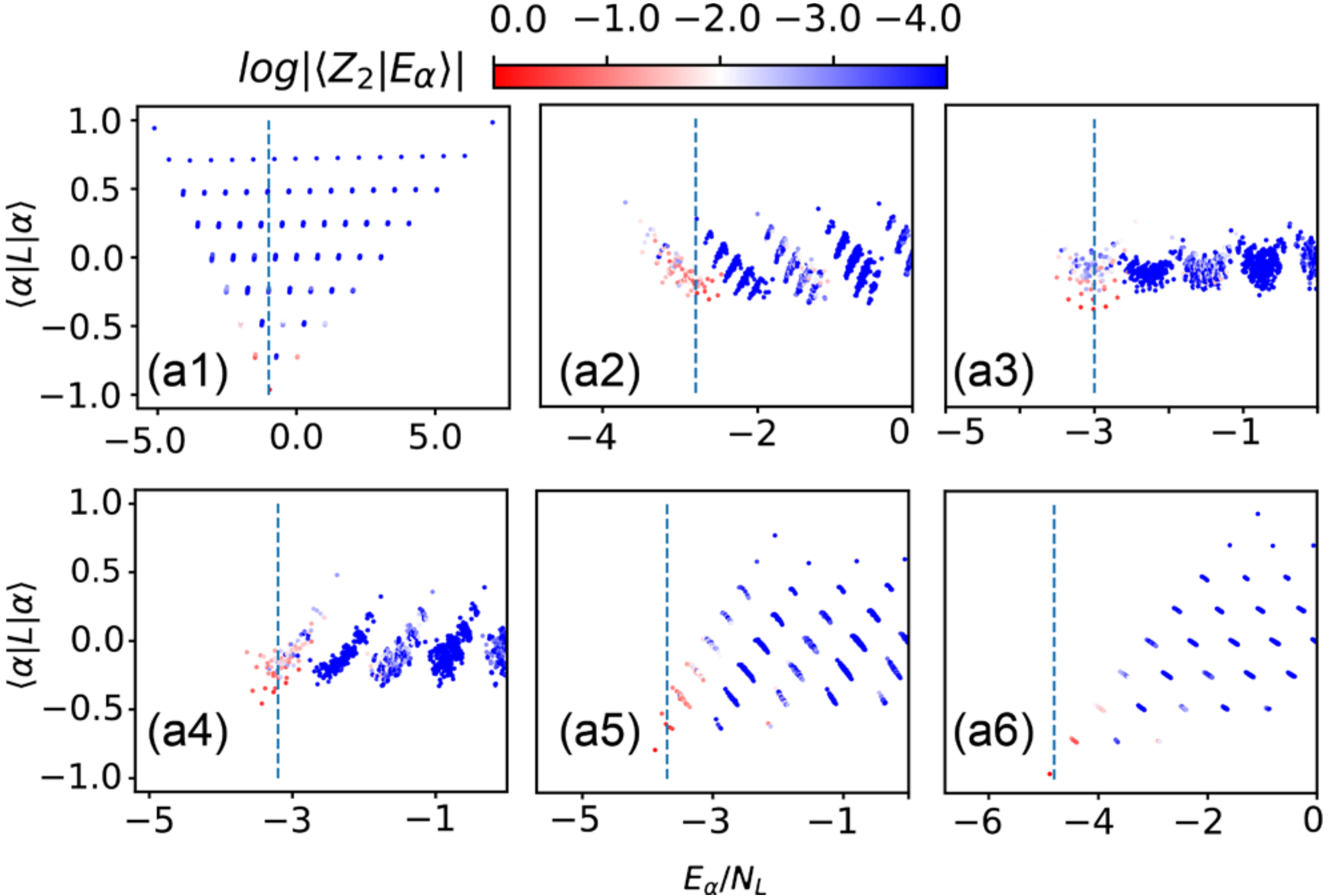}
\caption{Diagonal elements of $\hat{L}$ in energy eigenbasis at a fixed $h=6$ and different $J=1$(a1), $2.8$(a2), $3(=h/2)$(a3), $3.2$(a4), $3.7$(a5), $4.8$(a6). The lattice size is $N_L=16$. In all plots, the color shows the overlap between each eigenstate and initial $|Z_2\rangle$ state, and the vertical dashed line marks the location of initial state energy $E(t=0)$. }
\label{crossover}
\end{figure}

The first crossover is from single-particle limit (small $J$) to QMBS ($J=h/2$), with the  distributions of $\langle L \rangle_{\alpha}$ shown in Fig.\ref{crossover}(a1,a2,a3). In the single-particle limit $J\rightarrow 0$, the system is exactly solvable with well separated energy levels. As shown in Fig.\ref{crossover}(a1), since the initial energy of $|Z_2\rangle$ lies in the middle of the spectrum with a wide $\langle L \rangle_{\alpha}$ distribution, one can anticipate a non-thermal dynamics starting from this state. In the QMBS point with $J=h/2$, the system is also non-thermal but with a different mechanism, as analyzed previously and shown in Fig.\ref{crossover}(a3). For an intermediate $J$ between $\sim 0$ and $h/2$, the system is always non-thermal with a visibly finite $\Delta \bar{L}$, as seen from Fig.\ref{fig_QPT}(a3). The switch of non-thermal mechanism, from single-particle to QMBS physics, roughly occurs at $J=2.8$ where $|\Delta \bar{L}|$ shows a minimum (Fig.\ref{fig_QPT}(a3)) and $\langle L \rangle_{\alpha}$ distribution is most converged (Fig.\ref{crossover}(a2)). Further increasing $J$ to beyond $h/2$, the system departs from QMBS and tends to be more thermal, as manifested by the less-ordered distribution of $\langle L \rangle_{\alpha}$ as shown in Fig.\ref{crossover}(a4).

The second crossover is from QPT (at $J_c$) to classical Ising limit (large $J$), with the  distributions of $\langle L \rangle_{\alpha}$ shown in Fig.\ref{fig_QPT2}(c2) and Fig.\ref{crossover}(a5,a6). As analyzed previously, QPT can enhance the quantum ergodicity starting from low-energy states, thereby leading to a converged $\langle L \rangle_{\alpha}$ distribution (Fig.\ref{fig_QPT2}(c2)) and a negligible $\Delta \bar{L}$ signifying thermalization. Increasing $J$ to be larger than $J_c$, the system departs from the QPT-enhanced thermalization and becomes non-thermal. This is shown in Fig.\ref{crossover}(a5), where $\langle L \rangle_{\alpha}$ distribution gets more expanded and meanwhile  the initial energy of $|Z_2\rangle$ moves closer to the ground state energy. In the limit of classical Ising chain with large $J$, as shown in Fig.\ref{crossover}(a6), $|Z_2\rangle$ can well approximate the ground sate such that the system becomes thermal again. This is very similar to the situation discussed previously with zero $h$ and large $J$.

From above, we can see that the first crossover connects two non-thermal regimes with distinct mechanisms, while the second crossover connects two thermal regimes with distinct mechanisms. Fig.\ref{crossover} shows how the system evolves during each crossover, from which a general rule can be concluded. Namely, during the crossover between distinct thermalization (or non-thermalization) regimes in parameter space, the system tends to behave more non-thermal (or thermal).  In fact, the same rule also applies to the non-thermal crossover of Rydberg system with tunable $V$, where the system tends to be more ergodic (with small $|\Delta \bar{L}|$) for intermediate values of $V$, 
see Fig.\ref{fig_finite_V} and Fig.\ref{fig_finite_V_2}.

\section{Thermalization property starting from $|Z_0\rangle$} \label{sec_IV}

In this section, we consider the dynamical evolution of ICTLF system starting from a different initial state, i.e., the ferromagnetic state $|Z_0\rangle=|\downarrow \downarrow ...\downarrow\downarrow\rangle$. The according  phase diagram in the ($h,J$) plane is presented in Fig.\ref{diagram}(b). We can see that this diagram is substantially different from  Fig.\ref{diagram}(a), the diagram for $|Z_2\rangle$ initial state. To highlight the difference, we will also discuss the $|Z_0\rangle$-initiated dynamics along the line $J/h=1/2$, which describes the Rydberg atoms with tunable nearest-neighbor interaction. Moreover, another important feature of Fig.\ref{diagram}(b) is the emergence of thermalization {\it belts}  at some special ratios of $J/h$, see gray dash-dot lines therein. These thermalization belts can be attributed to the resonant spin flip on top of $|Z_0\rangle$ initial state, as we will explain in this section.

\subsection{Non-thermal to thermal crossover along $J/h=1/2$}

As shown by Eq.(\ref{H_V}), the Rydberg Hamiltonian is exactly given by the line $J/h=1/2$ in ICTLF model, with Rydberg coupling $V\equiv4J=2h$. By continuously varying $V$ from zero to large, we will show that the $|Z_0\rangle$-initiated dynamical system undergoes a smooth non-thermal to thermal crossover. 

\begin{figure}[h]
\includegraphics[width=8cm]{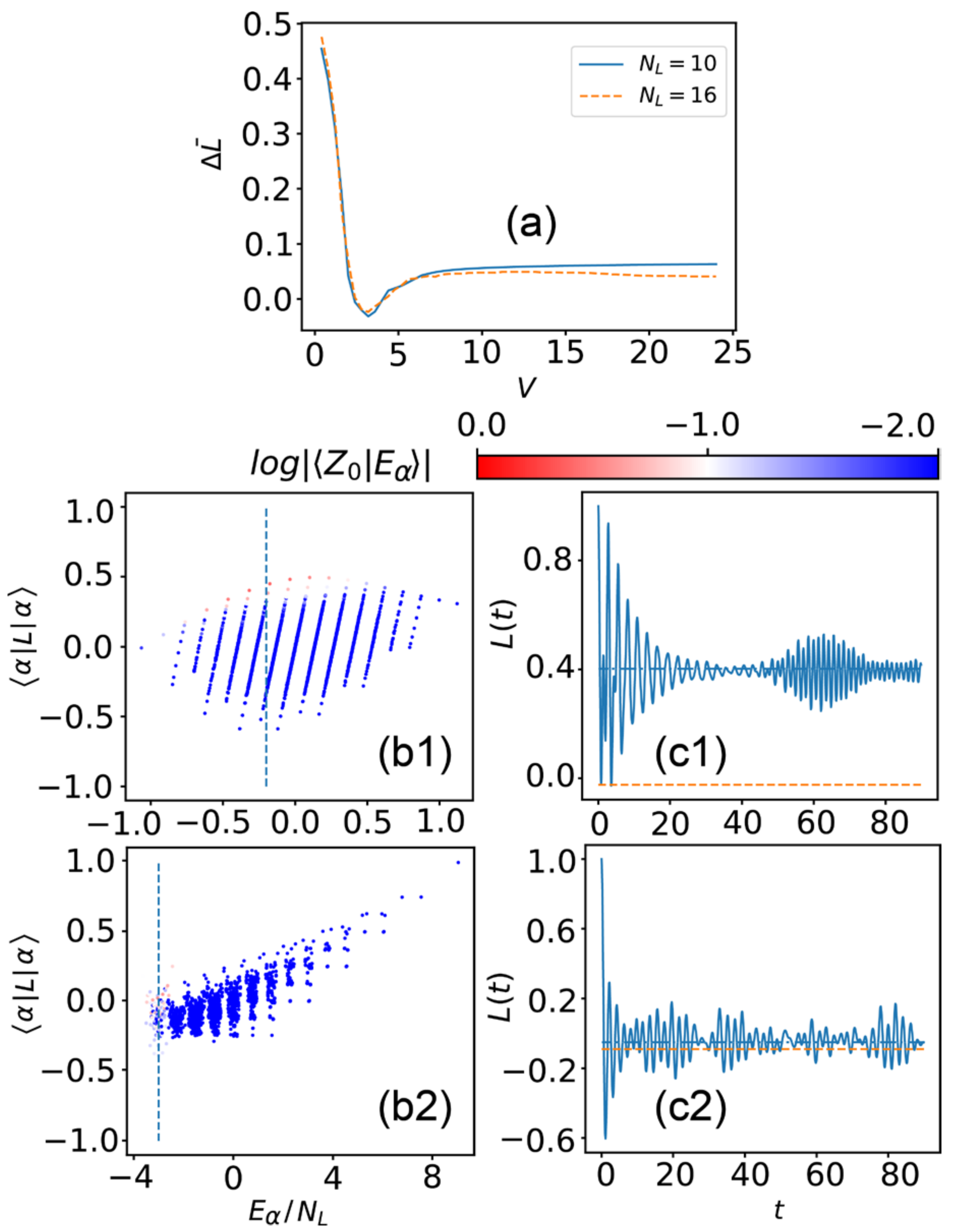}
\caption{Non-thermal to thermal crossover for Rydberg system with tunable interaction $V$ and starting from $|Z_0\rangle$ state. (a)  $\Delta \bar{L}$ as a function of  $V$ at lattice sizes $N_L=10,16$. 
(b1,b2)Diagonal elements of $\hat{L}$ in energy eigenbasis for $V=0.8$ (b1) and $V=12$ (b2). The color shows the overlap between each eigenstate and initial $|Z_0\rangle$ state, and the vertical dashed line marks the location of initial state energy $E(t=0)$. (c1,c2) Dynamics $L(t)$ for $V=0.8$ (c1) and $V=12$ (c2). The red dashed line shows the thermal prediction $\bar{L}_{th}$, and the blue dot-dashed  line in  shows $\bar{L}$. Here $N_L=16$ for (b1,b2,c1,c2). }
\label{fig_Z0_Ryd}
\end{figure}

In Fig.\ref{fig_Z0_Ryd}(a), we show $\Delta \bar{L}$ as a function of Rydberg coupling $V$ for different lattice sizes $N_L$. One can see that as increasing $V$ from zero, $\Delta \bar{L}$ decrease rapidly from a large value, indicating the system quickly evolving outside the non-thermal QI regime. At sufficiently large $V$, $\Delta \bar{L}$ stays at a very small value, and this value becomes smaller for larger $N_L$. This is consistent with the thermalization of $|Z_0\rangle$-initiated dynamics at large Rydberg coupling\cite{Lukin2017Rydberg}. In Fig.\ref{fig_Z0_Ryd}(b1,c1) and (b2,c2), we take two typical values of $V$ and examine the diagonal elements of $\hat{L}$ in energy space as well as its actual dynamics as time goes. One can see that for small $V$ near the QI regime, the $\langle L \rangle_{\alpha}\sim E_{\alpha}$ distribution in Fig.\ref{fig_Z0_Ryd}(b1) exhibit clear stripes, and as a result the long-time dynamics of $L(t)$ cannot be predicted by the thermal value, as shown in Fig.\ref{fig_Z0_Ryd}(c1).  In comparison, for large  $V$ in Fig.\ref{fig_Z0_Ryd}(b2), the stripes smear out, and there are a large amount of states nearby the initial energy. Different from the case of  $|Z_2\rangle$ initial state(Fig.\ref{fig_finite_V_2}(a3)), the states that have largest overlap with $|Z_0\rangle$ mostly distribute within the energy continuum. Consequently, $L(t)$ tends to thermalize at long time, as shown by Fig.\ref{fig_Z0_Ryd}(c2).

\begin{figure}[h]
\includegraphics[width=6cm, height=8cm]{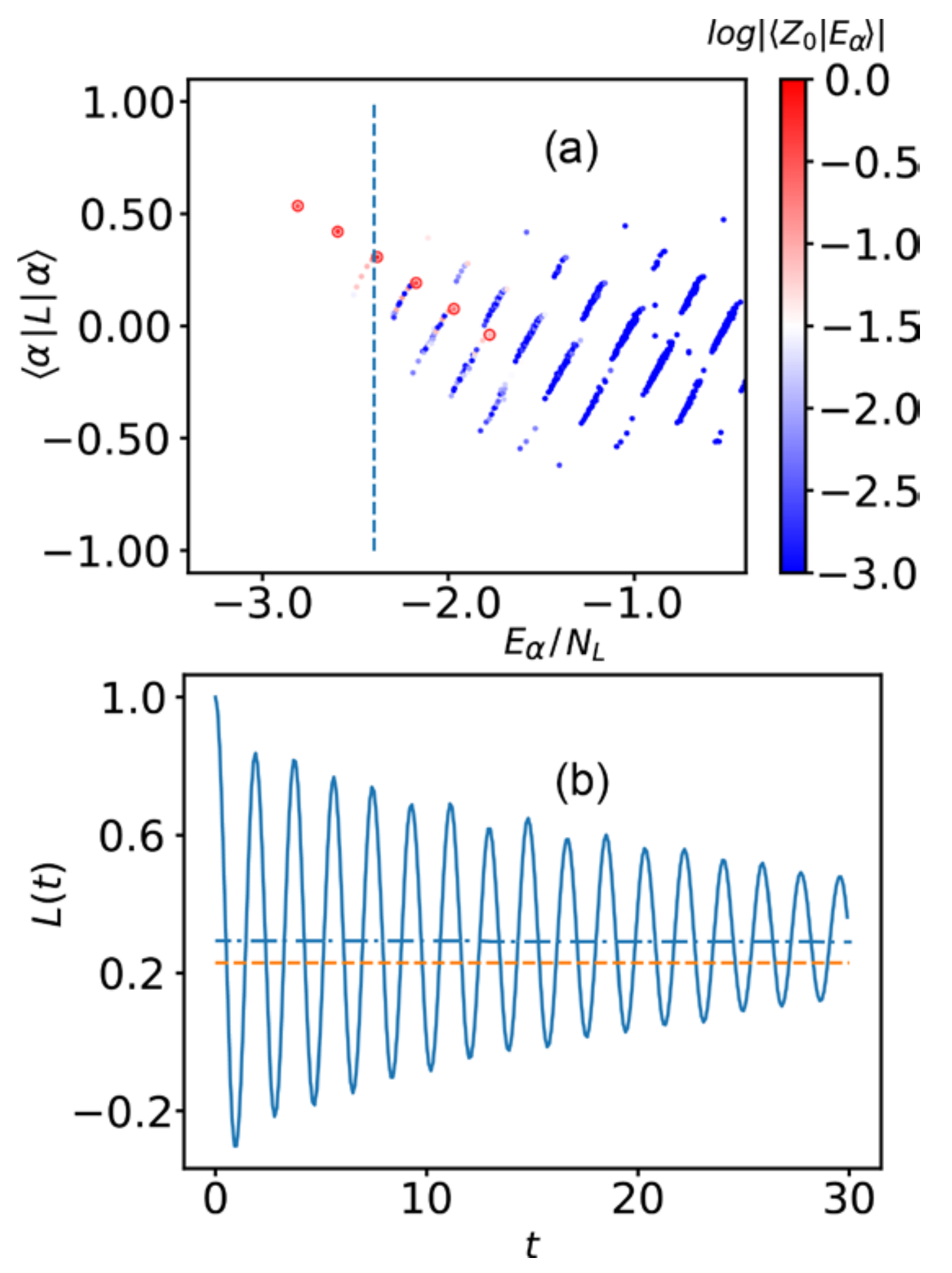}
\caption{Quantum many-body scar for the $|Z_0\rangle$-initiated dynamics at $h=4,\ J=1.6$. (a)Diagonal elements of $\hat{L}$ in energy eigenbasis. The color shows the overlap between each eigenstate and initial $|Z_0\rangle$ state, and the vertical dashed line marks the location of initial energy $E(t=0)$. The scar states are highlighted by red circles. (b) Actual dynamics of $L(t)$. The red dashed line shows the thermal prediction $\bar{L}_{th}$, and the blue dot-dashed  line in  shows $\bar{L}$. Here $N_L=16$.}
\label{fig_Z0_QMBS}
\end{figure}

Here we would also like to mention that in the limit $J,\ h\rightarrow\infty$ but below the line $J/h=1/2$, our system reproduce the PXP Hamiltonian with an additional magnetic field $h_{\rm eff}\equiv h-2J>0$, which was shown to equally support QMBS starting from $|Z_0\rangle$ state\cite{Yuan2022Z0scar}. We have confirmed this in Fig.\ref{fig_Z0_QMBS} by taking $h=4$ and $J=1.6<h/2$, which gives $h_{\rm eff}=0.8$. It shows that there exists a sequence of scar states with equal  energy spacing that have largest overlap with initial $|Z_0\rangle$ state, as highlighted by red circles in Fig.\ref{fig_Z0_QMBS}(a). As a result, the time evolution of $L(t)$ shows visible periodic oscillation at long time and does not thermalize, see Fig.\ref{fig_Z0_QMBS}(b). These are all typical features of non-thermal dynamics caused by QMBS. 

\subsection{Thermalization facilitated by resonant spin flipping}

\begin{figure}[t]
\includegraphics[width=8.5cm]{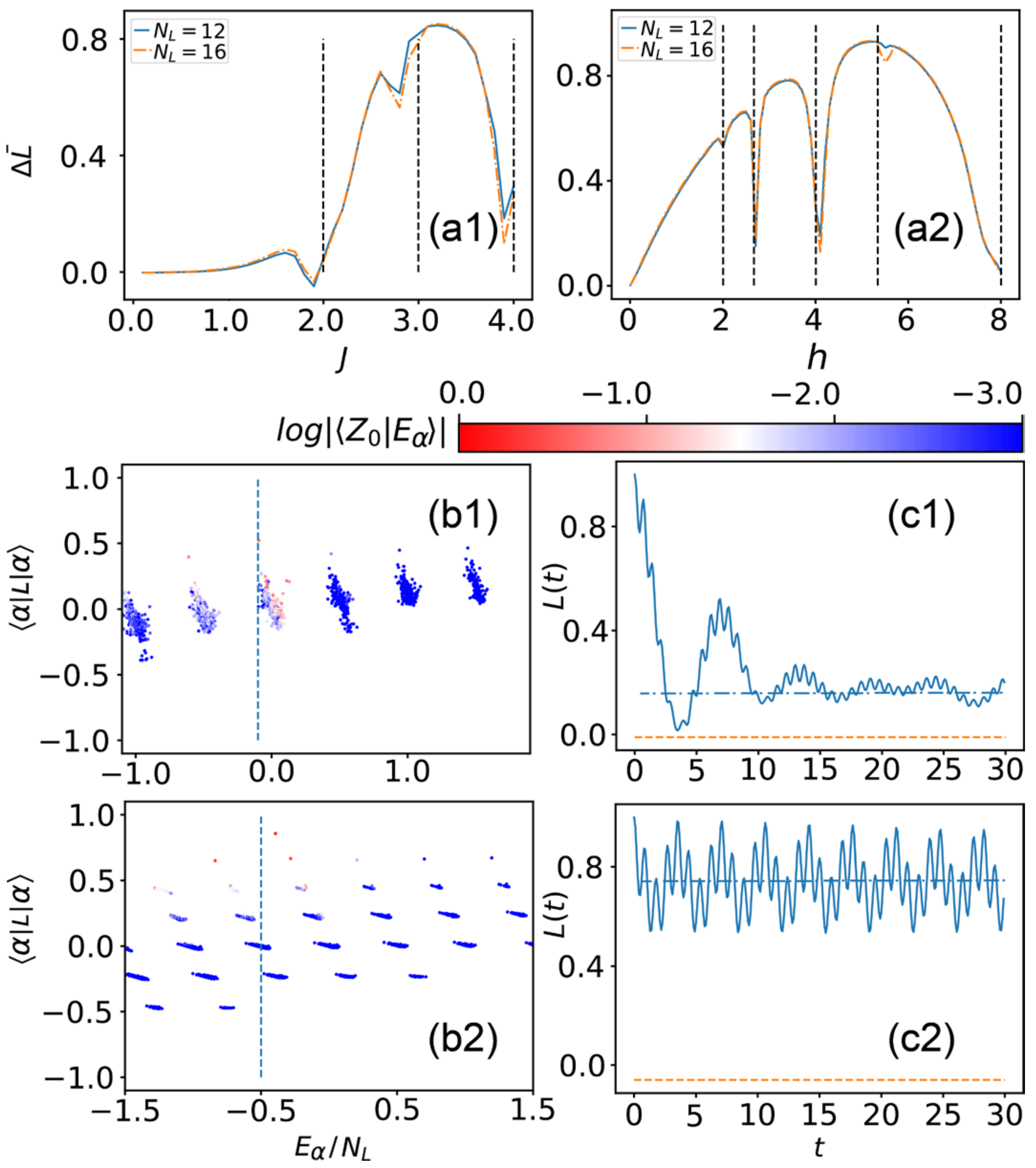}
\caption{ Thermalization enhanced by resonant spin flipping on top of initial $|Z_0\rangle$ state. (a1) $\Delta \bar{L}$ as a function of $J$ for fixed $h=4$. The vertical lines denote the special ratios at $J/h=1/2,\ 3/4$ and $1$ (from left to right), which are close to the minimum of $\Delta \bar{L}$. (a2) $\Delta \bar{L}$ as a function of $h$ for fixed $J=4$. The vertical lines denote the special ratios $J/h=2,\ 3/2,\ 1,\ 3/4,\ 1/2$(from left to right) which are close to the local minima.
(b1,b2)Diagonal elements of $\hat{L}$ in energy eigenbasis for $h=4$ and $J=3.9(\sim h)$ (b1),  $3.5$ (b2). The color shows the overlap between each eigenstate and initial $|Z_0\rangle$ state, and the vertical dashed line marks the location of initial energy $E(t=0)$. (c1,c2) Dynamics of $L(t)$ corresponding to (b1,b2). The red dashed line shows the thermal prediction $\bar{L}_{th}$, and the blue dot-dashed  line in  shows $\bar{L}$.  Here $N_L=16$ for (b1,b2,c1,c2).}
\label{fig_Z0_resonance}
\end{figure}

Another important feature shown by Fig.\ref{diagram} (b) is the existence of several straight belts (white color with $\Delta \bar{L}\sim 0$) that extend to large $J,h$ regime, where the dynamical system tends to thermalize.  These belts are actually associated with some special ratios of $J/h$. To see it clearly, in Fig.\ref{fig_Z0_resonance}(a1,a2) we plot out $\Delta \bar{L}$ as a function of $J$ for a fixed $h=4$,   or as a function of $h$ for a fixed $J=4$.  We can see that $\Delta \bar{L}$ drops sharply near some special ratios: 
\begin{equation}
\frac{J}{h}=\frac{1}{2}, \ \frac{3}{4}, \ 1,\ \frac{3}{2},\ 2, ...  \label{ratio}
\end{equation}
As explained below, these special ratios can greatly facilitate thermalization by supporting resonant spin flips on top of $|Z_0\rangle$. 

\begin{figure}[h]
\includegraphics[width=8cm]{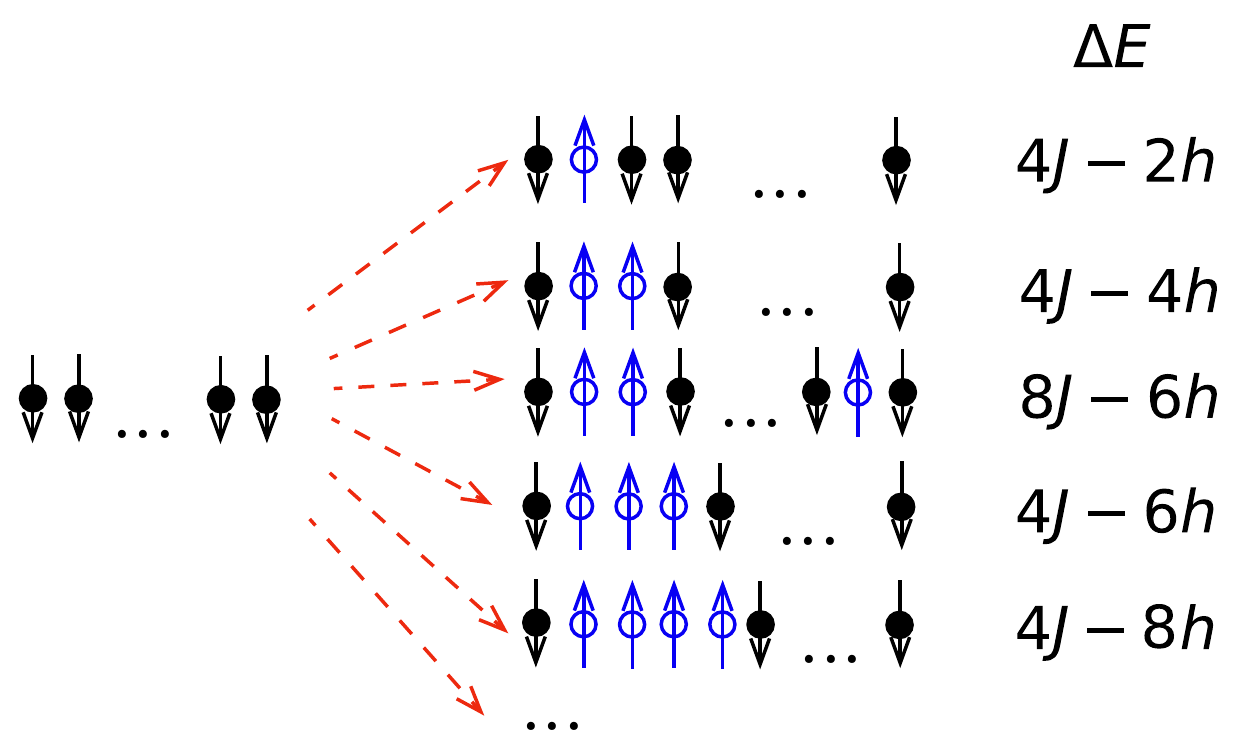}
\caption{Schematics for the resonant spin flip starting from $|Z_0\rangle$ (left state). In the limit of large $J$ and $h$, the system becomes a classical Ising chain. At the right side, we list several typical configurations of spin flip and its according energy change $\Delta E$. The resonant spin flip occurs when $\Delta E=0$, giving rise to special ratios of $J/h$ as listed in (\ref{ratio}). }
\label{fig_Z0_schematic}
\end{figure}

The resonant spin flip can be understood straightforwardly in large $J,h$ limit, where the system corresponds to a classical Ising chain and one can estimate the energy easily.  In Fig.\ref{fig_Z0_schematic}, we list several typical configurations of spin flip on top of $|Z_0\rangle$, together with the according energy change after the flip ($\Delta E$). When the ratio $J/h$ takes special values as listed in (\ref{ratio}), there exists a sequence of similar configurations with zero energy cost ($\Delta E=0$), which can be referred as resonant spin flips.  For instance, the resonant one spin flip occurs at $J/h=1/2$, and resonant two nearest-neighbor-flip occurs at $J/h=1$, etc. There can also be combination of one flip at some place and two flips at other place, which becomes resonant with $|Z_0\rangle$ when $J/h=3/4$. These resonant spin flips lead to more accumulation of states near the initial energy which can also have  large wave-function overlap with $|Z_0\rangle$,  as shown in Fig.\ref{fig_Z0_resonance} (b1) for the case of  $J/h\sim 1$. This greatly enhances the quantum ergodicity and drive the system to approach thermalization. 
In comparison, when depart from the special ratios, as shown in    
Fig.\ref{fig_Z0_resonance} (b2,c2), there will be rare states having similar energy or  large overlap with $|Z_0\rangle$ (Fig.\ref{fig_Z0_resonance} (b2)), and therefore the resulted dynamics is far away from the thermalization regime, as manifested by much larger long-time difference $\Delta \bar{L}$ (Fig.\ref{fig_Z0_resonance}(c2)) than the resonant case (Fig.\ref{fig_Z0_resonance}(c1)).

\section{Summary and discussion}\label{sec_V}

In summary, we have studied the thermalization property of ICTLF model starting from the anti-ferromagnetic ($|Z_2\rangle$) or  ferromagnetic ($|Z_0\rangle$) initial states.  Specifically, such property is investigated by comparing the real-time dynamics of local observable with its thermal predictions. Here, a large deviation between them is a clear evidence for non-thermalization of the system. The resulted phase diagrams in the parameter plane of Ising coupling($J$) and longitudinal field($h$), as shown by Fig.\ref{diagram}(a) and (b), are dramatically different. Starting from $|Z_2\rangle$, we have revealed a non-thermal crossover from QI to QMBS by following the line with $J/h=1/2$, which corresponds to the Rydberg Hamiltonian with tunable nearest-neighbor interactions. This gives an intriguingly first example when the two distinct ETH-breaking mechanisms can smoothly connect to each other in the parameter space. In contrast, starting from $|Z_0\rangle$, we observe a non-thermal to thermal crossover by following the same line. All these phenomena can be directly probed in Rydberg systems with tunable interactions.

Furthermore, in this work we have demonstrated two underlying mechanisms for thermalization. One is the quantum phase transition(QPT), as located by the gray dashed line in the diagram for $|Z_2\rangle$ initial state (Fig.\ref{diagram}(a)); the other is the resonant spin flip, as shown by a number of thermalization belts (gray dash-dot lines) in the diagram for $|Z_0\rangle$ initial state (Fig.\ref{diagram}(b)). We emphasize that the two mechanisms for thermalization all strongly rely on the initial states chosen. For instance, the QPT-facilitated thermalization requires the initial state lying in the low-energy manifold, such that the quantum ergodicity can be achieved taking advantage of the large correlation length of low-energy states near a QPT. This may offer a new way to characterize QPT, i.e.,  through the thermalization of dynamical system starting from low-energy states. For the second mechanism,  it also strongly depends on the initial state, which provides the substrate for  spin flip and determines the location of  thermalization belts in parameter space. All these results could be detected using interacting bosons under a tilted lattice\cite{Greiner2011IsingExps, Yuan2022Z0scar}.  Given the robust physics of above two mechanisms, they are expected to apply for the thermalization phenomena in a wide class of interaction models and physical systems.

{\it Acknowledgements.} 
We thank Hui Zhai, Lei Pan and Zhiyuan Yao for helpful discussions. The work is supported by the National Key Research and Development Program of China (2018YFA0307600), the National Natural Science Foundation of China (12074419, 12134015), and the Strategic Priority Research Program of Chinese Academy of Sciences (XDB33000000).


\appendix

\section{Comparison of phase diagrams when choosing another momentum intervals in the thermal average} \label{appendix_a}

In the main text we have chosen the momentum interval as $K\in[0,2\pi)$ in computing the thermal average. This may cause a problem for the dynamics is initiated from $|Z_0\rangle$, which has total zero momentum and thus only involves $K=0$ subspace. Because of this, it is necessary to confirm the result by choosing a different momentum interval to compute $\bar{L}_{th}$, such that all momentum states can be involved in the numerical calculations. Here we will take the interval $K\in(-\pi,\pi]$ as a typical example.

\begin{figure}[h]
\includegraphics[height=7cm]{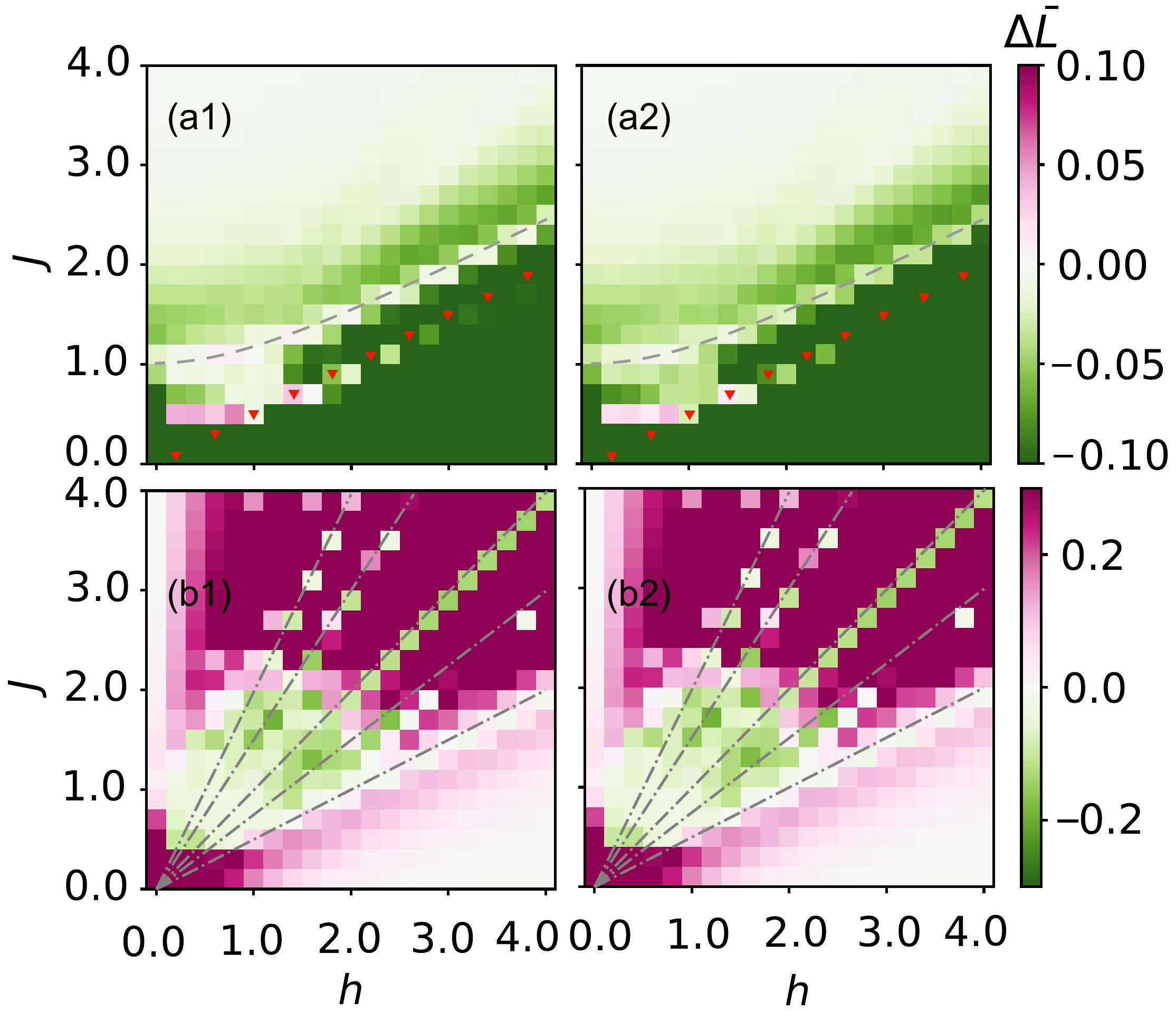}
\caption{Dynamical phase diagrams starting from $|Z_2\rangle$(a1,a2) or $|Z_0\rangle$(b1,b2) state when choosing different momentum intervals in the thermal average. The momentum interval is $[0,2\pi)$ in (a1,b1) and $(-\pi,\pi]$ in (a2,b2). The results are obtained after the finite-size scaling, i.e., for an infinite chain ($N_L=\infty$). }\label{app_Linf-K_interval_comp}
\end{figure}

\begin{figure}[h]
\includegraphics[height=7cm]{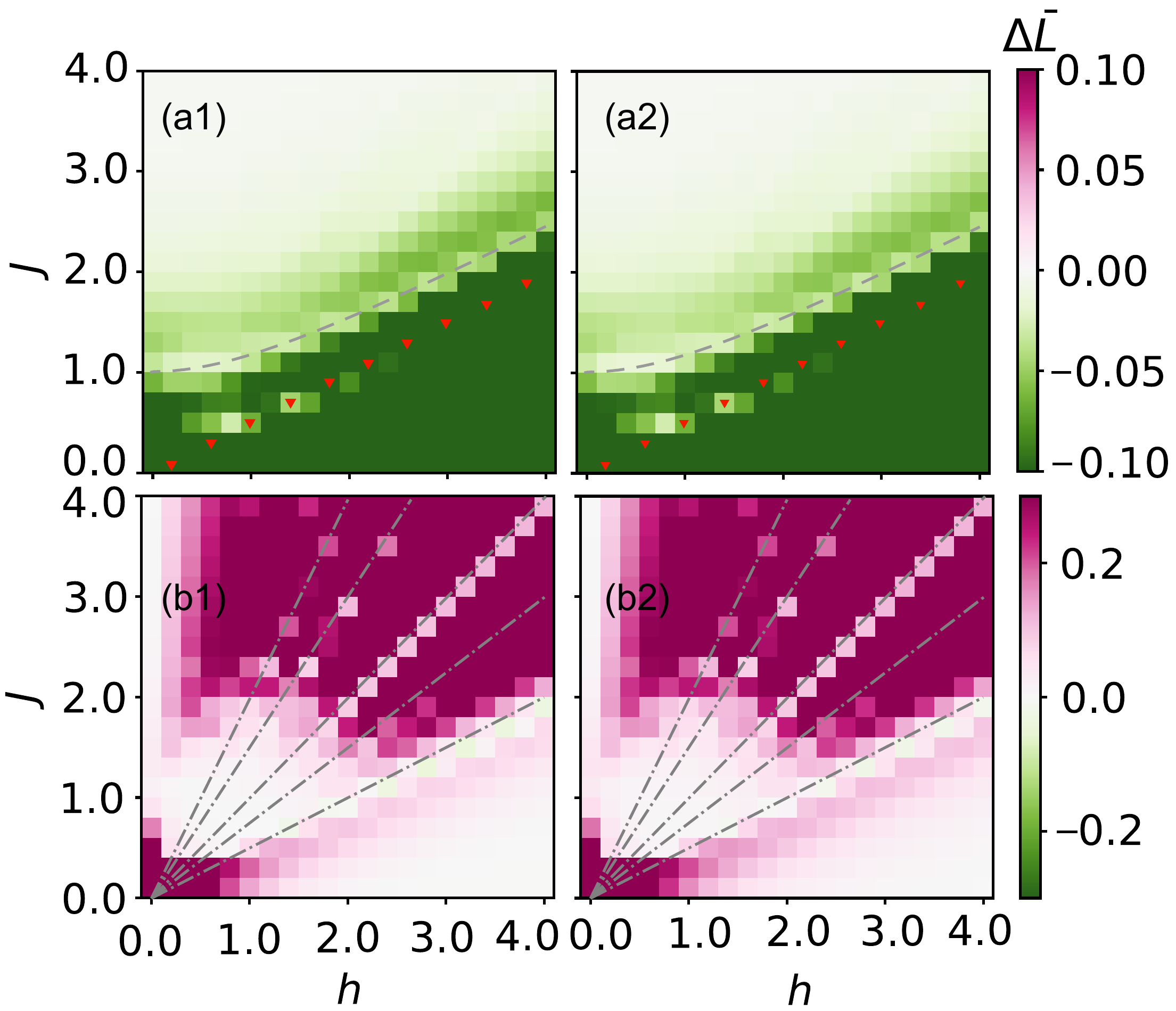}
\caption{The same as Fig.\ref{app_Linf-K_interval_comp} except for a finite lattice size $N_L=16$.}\label{app_L16-K_interval_comp}
\end{figure}

In Fig.\ref{app_Linf-K_interval_comp}, we show the $|Z_2\rangle$- and $|Z_0\rangle$-initiated phase diagrams for an infinite chain by choosing different momentum intervals to compute the thermal average. Fig.\ref{app_Linf-K_interval_comp}(a1,b1) are for $K\in[0,2\pi)$, the same as Fig.1 in the main text. In comparison, Fig.\ref{app_Linf-K_interval_comp}(a2,b2) are for $K\in(-\pi,\pi]$. Due to the numerical cost we only show a sparser grid of $\Delta h=\Delta J=0.2$. One can see that the two groups of diagrams show consistent key features, including the non-thermal crossover for the Rydberg model and the quantum ergodicity enhanced by quantum phase transition and spin flip. To eliminate the effect of finite-size scaling, in Fig.\ref{app_L16-K_interval_comp} we further compare the phase diagrams for a finite lattice size $N_L=16$, and again find good consistence therein.

\section{Comparison of phase diagrams between infinite and finite lattice sizes} \label{appendix_b}

 Fig.\ref{app_Linf-K_interval_comp} and Fig.\ref{app_L16-K_interval_comp} also give the information for the comparison between an infinite chain (after finite-size scaling) and a finite chain (with  $N_L=16$). 
Take a fixed momentum interval $[0,2\pi)$ for example,   one can see that Fig.\ref{app_Linf-K_interval_comp}(a1,b1) and Fig.\ref{app_L16-K_interval_comp}(a1,b1)  are qualitatively consistent with each other. Namely, they both well reflect the key features of the system, including the non-thermal crossover for the Rydberg model and the quantum ergodicity enhanced by quantum phase transition and spin flip. Therefore we can conclude that these features are very robust that do not rely on the size of the system. Meanwhile, we also see that the features after finite-size scaling, as shown in Fig.\ref{app_Linf-K_interval_comp}(a1,b1), are more visible than the finite-size case shown in Fig.\ref{app_L16-K_interval_comp}(a1,b1). This suggests the enhanced features  in larger systems. The same also applies to the case with a different momentum interval $(-\pi,\pi]$, see the comparison between Fig.\ref{app_Linf-K_interval_comp}(a2,b2) and Fig.\ref{app_L16-K_interval_comp}(a2,b2).




\end{document}